\documentclass[aps,prb,preprint,showpacs,superscriptaddress,footnoteinbib]{revtex4}
\usepackage{latexsym}
\usepackage{graphicx,epsfig}
\usepackage{amssymb, amsmath}
\begin{document}
\title{Charge and spin density waves: Quasi one dimension to two  dimensions}
\author{Urbashi Satpathi}
\address{Unit for Nano Science and Technology,
S. N. Bose National Centre for Basic Sciences, JD Block,
Sector III, Salt Lake City, Kolkata 98, India.}
\author{Sumit Ghosh}
\address{Unit for Nano Science and Technology,
S. N. Bose National Centre for Basic Sciences, JD Block,
Sector III, Salt Lake City, Kolkata 98, India.}
\author{Ashim Kumar Ray}
\address{Physics and Applied Mathematics Unit, 
Indian Statistical Institute, 
203 Barrackpore Trunk Road, Kolkata 108, India.}
\author{P. Singha Deo}
\address{Unit for Nano Science and Technology,
S. N. Bose National Centre for Basic Sciences, JD Block,
Sector III, Salt Lake City, Kolkata 98, India.}
\begin{abstract}
Electronic charge and spin separation leading to charge density 
wave and spin density wave is well established in one dimensional 
systems in  the presence and absence of Coulomb interaction. We start from 
quasi one dimension and show the possibility of such a transition in 
quasi one dimension as well as in two dimensions by going to a regime 
where it can be shown for free electrons that just interact via Fermi 
statistics. Since Coulomb interaction can only facilitate the 
phenomenon, the purpose of our work is to show the phenomena 
unambiguously in the limit when Coulomb interaction can be ignored. 
Finally we also comment on dimensions greater than two and including 
Coulomb interactions.
\end{abstract}
\maketitle

\section{Introduction}
Linear superposition principle in quantum mechanics tells us that spontaneous symmetry breaking is not possible in quantum mechanics as the infinite number of degenerate states that are associated with spontaneous symmetry breaking can superpose to give a general state that has the same symmetry as the Hamiltonian. 
In spite of it, certain heavy nuclei exhibit rotational excitations, 
that can not be explained by the shell structure alone of a spherical nucleus. 
Initial understanding of this was provided by Bohr and Mottelson \cite{Bohr} 
in terms of collective modes of oscillation of a deformed nucleus. Such nuclear 
deformation may well be due to spontaneous symmetry breaking.  
However as nuclear forces and nuclear Hamiltonian are still not precisely 
known, a first principle quantum mechanical analytic understanding is not yet possible. 
Similar ideas of spontaneous symmetry breaking can also explain the details 
of the mass spectra of alkali metal clusters and indicate the existence 
of a spin-density wave in quantum dots \cite{RM,KOS,RMP,KOSK}.

In these systems, since earlier days, researchers have approached 
the problem from two practical points of view although it becomes difficult to 
obtain a clear, reconciled understanding. In the first approach, one either makes a 
numerical solution of Hartree-Fock equations or the Kohn-Sham equations for a few electrons and obtain the electron density. The density profile shows a typical 
crystal like structure consisting of hills and 
valleys \cite{RM,KOS,RMP}. It is known that this is an artifact of the non 
linearity of the equations in use while the exact theory is linear 
\cite{VIEFERS}. In the other approach, one makes an exact diagonalization for even fewer electrons. Whereby, the density do not show any signature of 
broken symmetry because of the linear superposition principle. 
However if one calculates the pair correlation function, then that shows 
oscillations \cite{RMP,PHYSICA,FILI}. Generally speaking, these oscillations survive over a finite length and decay rapidly which is expected in finite size 
systems. The correlation function is not defined in Hartree-Fock Theory or 
Density Functional Theory. The density oscillations obtained therein do 
not decay beyond a length scale. Still one makes the ad hoc assumption 
that the non linearity of Hartree-Fock Theory or Density Functional 
Theory show density oscillations by projecting the pair correlation 
function and consequently, the discrepancy of the decay disappears for infinite 
systems. With exact diagonalization, one also looks at the degeneracy 
of eigen energies after subtracting the center of mass energy and checks 
if the degeneracy can be explained by the representative point group corresponding 
to the broken symmetry crystalline state \cite{RMP}. Once again, one 
cannot go to very large energy limits due to numerical problems. Already at higher energies 
one starts to notice deviations from exact degeneracies just as oscillations in 
pair correlation function decay beyond a certain length scale \cite{KOSK}.
 
\begin{figure}
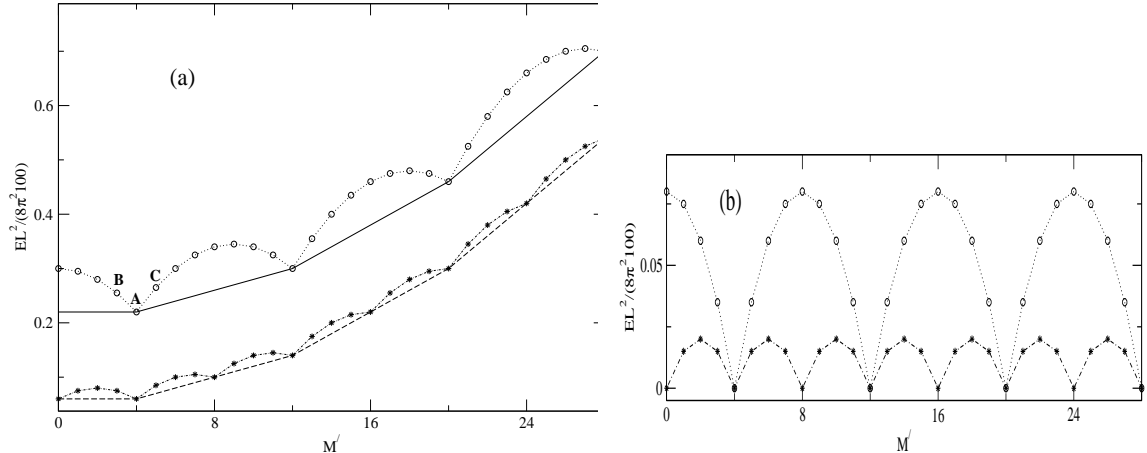

\includegraphics[height=6truecm,width=8truecm]{1d.eps}
\includegraphics[height=4truecm,width=7truecm]{1db.eps}
\caption{(a) Yrast spectra for electrons in a one dimensional ring. 
The electrons are interacting only through statistics whence Coulomb 
interaction is ignored. 
The circles and the stars are the calculated data points. The circles are 
for 8 spin up electrons (connected by dotted line) and the stars are 
for 4 spin up and 4 spin down electrons (connected by dash-dot line). 
The solid line is drawn by connecting the local minima of the 8 spin 
up electrons and dashed line is drawn connecting the local minima for 
the 4 spin up 4 spin down  electrons. The solid and dashed line fall on a 
parabola given by $M^2/2I$ where M is the total angular momenta and $ I $ is 
the moment of inertia for 8 classical particles in a one dimensional ring. 
(b) The energy values are plotted after subtracting solid and dashed 
line in Fig. \ref{1d}$a$ from dotted and dash-dotted lines,
respectively, in Fig. \ref{1d}$a$.}
\label{1d}
\end{figure} 

Mesoscopic systems give a unique opportunity to study the few electron 
system both experimentally as well as with theoretical models and hence provide 
an opportunity to study how few electron properties evolve into 
macroscopic collective properties as we increase the number of electrons. 
Wigner crystallization of electrons, one such bulk phenomenon of 
spontaneous symmetry breaking proposed long ago, is still a debatable 
issue. We exclude here the situation when quantum mechanical kinetic energy or the 
uncertainty of an electron can be quenched by a strong magnetic 
field \cite{RA,sc,FP,m13} or the situation when explicit symmetry 
breaking leads to an electron crystal state \cite{SZ,be,re,gu,SZ2}.

\section{One dimensional ring}
In one dimension, correlation due to Pauli exclusion principle alone can 
cause oscillation in pair correlation function within short distances 
even in absence of Coulomb interaction \cite{PHYSICA}. One dimensional 
system however, is an idealization that is well understood. If one takes a finite 
number of electrons in a one dimensional ring, the eigen 
energy can be calculated very easily. A typical curve is shown in 
Fig. \ref{1d}. The structure obtained in Fig. \ref{1d} 
can be understood in terms of spontaneous symmetry breaking \cite{PHYSICA}. 
The electrons form a crystal in the center of mass frame while the center of mass behaves 
like a free particle and hence the solid and dashed curve in Fig. \ref{1d} increases 
parabolically. The points B and C correspond to excitations that are 
therefore decoupled from the center of mass motion and corresponds to 
vibrations of localized electrons in center of mass frame. If the 
parabolic contribution is subtracted then the yrast spectra show periodic 
oscillation (Fig. \ref{1d}b). That is, if one subtracts the solid line from 
circular dots in Fig. \ref{1d}a,  one gets the dotted line in Fig. \ref{1d}b 
and if one subtracts the dashed line from the stars in Fig. \ref{1d}a, one 
gets the dash-dot line in Fig. \ref{1d}b. Figs. \ref{1d}a and \ref{1d}b imply 
that eight spin up electrons, crystallized in a one dimensional 
ring has eight-fold discrete symmetry while four up and four down has a 
four-fold discrete symmetry. Hence the yrast spectrum repeats modulo 
eight and four respectively in Fig. \ref{1d}b.  
As the length of the ring tends to infinity the points B and C in 
Fig. \ref{1d}a will come closer to the point A and we get Fig. \ref{1d} 
of Ref. [\onlinecite{halden}]. Such an infinite one dimensional system can be 
bosonized wherein the bosonic excitations are 
the above mentioned vibrations of localized charge that act as phonons. 
This is well studied using Luttinger liquid, Calogero-Sutherland model 
and Bethe Ansatz \cite{Mahan}. Two dimensional systems are not very well 
understood although there have been lot of efforts to show the same 
features in two dimensions due to the experimental observations in 
heavy fermion systems and high temperature superconductors 
\cite{Amato,Guy}. In this paper, we wish to show that quasi one dimensional system can 
provide very interesting clue to understand the corresponding higher dimensional analogues as well.

\section{Quasi one dimensional ring and two dimensions} 
Let us consider N number of electrons in a quasi one dimensional (Q1D) ring. 
The Hamiltonian in presence of a magnetic field perpendicular to the 
plane of the ring is given by
\begin{eqnarray}
H = \sum_{j=1}^N \left[\frac{1}{2m^*} \left( -i \hbar \nabla_j - \frac{e}{c}\overrightarrow{A}(\overrightarrow{r_j}) \right)^2 + V(r_j) \right] + \frac{1}{2}\frac{1}{4 \pi \epsilon} \sum_{i\neq j} \frac{e^2}{\vert \overrightarrow{r_i}-\overrightarrow{r_j}\vert}
\label{h}
\end{eqnarray}
where $V(r_j)$ is the confinement potential for jth electron defined as
\begin{eqnarray}
V(r_j) & = 0 ~ &for~ r_{in}\leq r_j \leq r_{out} \\
       & = \infty ~ &elsewhere \nonumber
\end{eqnarray}
Here $r_{in}$ is the inner radius and $r_{out}$ is the outer radius of 
the ring. We use a unit system where $\hbar$=1, $c$=1, $e$=1, 
$4\pi \epsilon_0$=1 and $m^*$=0.5 and we choose the Bohr 
radius ($R_B=\frac{4\pi \epsilon_0 \hbar^2}{m^* e^2} $) to be the unit 
of length. Such a Q1D ring can be experimentally realized \cite{dat}.

\subsection{ Single particle states}
Single particle Hamiltonian ($H_0$) in presence of a magnetic field is 
given by disregarding the last term in Eq. (\ref{h}) and the sum in the 
first term. Therefore, the index $j$ is dropped.
%\begin{equation}
%H_0 = \frac{1}{2m^*} \left( -i \hbar \nabla - \frac{e}{c}\overrightarrow{A} \right)^2 + V(r)
%\label{h0}
%\end{equation} 
The corresponding Schr{\"o}dinger equation is
\begin{eqnarray}
\left[ \frac{1}{2m^*} \left( -i \hbar \nabla - \frac{e}{c}\overrightarrow{A} \right)^2 + V(r) \right]\psi = E \psi
\label{s1}
\end{eqnarray}
We use two dimensional polar coordinates $ ( r, \theta) $.
The vector potential in Coulomb gauge ($\overrightarrow{\nabla}\cdot\overrightarrow{A}=0 $) is given by
\begin{eqnarray}
\overrightarrow{A}(\overrightarrow{r}) = \frac{B r_f^2}{2r}\hat{\theta} = \frac{\omega_c m^*c r_f^2}{2er} \hat{\theta}
\label{A}  
\end{eqnarray}
where $B$ is the magnetic field passing through a finite circle of radius $r_f<r_{in}$ and $\omega_c = (eB/m^*c)$ is the cyclotron frequency. The above equation can be solved \cite{Sekha1,Sekha2} and we outline below our solution as our boundary conditions are different.
Thus from Eq. (\ref{s1}) and Eq. (\ref{A}),
\begin{eqnarray}
\frac{1}{2m^*}\left[ -\hbar^2 \frac{1}{r}\frac{\partial }{\partial r} \left(r \frac{\partial }{\partial r}\right) - \hbar^2 \frac{1}{r^2} \frac{\partial^2}{\partial \theta^2} + i \left( \hbar m^* \omega_c r_f^2 \right) \frac{1}{r^2} \frac{\partial}{\partial \theta} + \left( \frac{m^* \omega_c r_f^2}{2} \right)^2 \frac{1}{r^2} + V(r)\right]\psi = E \psi
\label{sch1}
\end{eqnarray}
Let us consider the region within the ring ($ r_{in} \leq r \leq r_{out} $) where the potential $V(r)$ is zero. Multiplying both sides by $\frac{2m^*r^2}{\hbar^2}$, we get from Eq. (\ref{sch1}),
\begin{eqnarray}
\left[ r \frac{\partial }{\partial r} \left( r \frac{\partial }{\partial r}\right) + \frac{2m^*E}{\hbar^2}r^2 - \left( \frac{m^* \omega_c r_f^2}{2 \hbar} \right)^2 \right]\psi + \left[ - i  \frac{ m^* \omega_c r_f^2}{\hbar} \frac{\partial}{\partial \theta} + \frac{\partial^2}{\partial \theta^2} \right] \psi = 0
\label{sch2}
\end{eqnarray}
Eq. (\ref{sch2}) allows us to write
\begin{equation}
\psi (r,\theta)= R(r) \Theta (\theta)
\label{psi}
\end{equation}
Thus from Eq. (\ref{sch2}) and Eq. (\ref{psi}) we get
\begin{eqnarray}
& & r \frac{d}{dr} \left( r \frac{dR}{dr} \right) + \left(\frac{2m^*E}{\hbar^2}r^2 - \alpha^2 - \lambda \right)R=0  \label{r1} \\
& & \frac{d^2 \Theta}{d \theta^2} - i 2 \alpha \frac{d\Theta}{d\theta} + \lambda \Theta = 0 \label{p1}
\end{eqnarray}
where $\lambda$ is a separation constant and $\alpha$ is given by
\begin{equation}
\alpha = \frac{m^* \omega_c r_f^2}{2 \hbar} = \frac{\Phi}{\Phi_0}
\label{alpha}
\end{equation}
 $\Phi$ is the flux passing through the ring and $\Phi_0$ is the flux quantum given by $\frac{ch}{e}$.
Solution for Eq. (\ref{p1}) is given by
\begin{equation}
\Theta(\theta) = \frac{1}{\sqrt{2\pi}}exp\left[i \left( \alpha \pm \sqrt{\alpha^2+\lambda} \right) \theta \right]
\end{equation}
The azimuthal wave function $\Theta(\theta)$
satisfy twisted periodic boundary
condition, i.e.,  $\Theta(\theta)=\Theta(\theta +2\pi) e^{-i{2 \pi \phi
\over \phi_0}}$ that gives
\begin{equation}
\alpha \pm \sqrt{\alpha^2+\lambda} = \alpha \pm m' \hspace{2cm} 
\label{lambda}
\end{equation}
where, $ m'= \sqrt{ \alpha^2 + \lambda }= 0,\pm 1,\pm 2,\ldots $
%and
%\begin{equation}
%m' = \pm \sqrt{ \alpha^2 + \lambda }
%\end{equation}

\begin{figure}                              
\includegraphics[height=7truecm,width=12truecm]{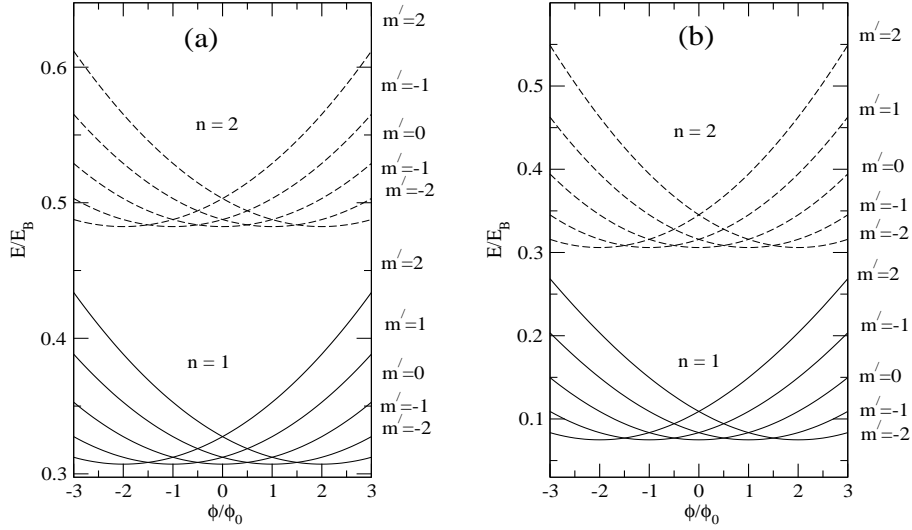}
\caption{Single particle eigen energies vs magnetic flux for an electron 
in a Q1D ring corresponding to 
$m'$ = -2 to $m'$ = 2 for $n$ = 1 and $n$=2 where the energy is expressed in 
units of $E_B=\frac{\hbar^2}{m^*R_B^2}$.
In (a) the inner radius is 8$R_B$  and the outer radius is 12$R_B$. In
(b) the inner radius is 4$R_B$ and the outer radius is 12$R_B$. In (a), we have subtracted 
0.75$E_B$ from the energy value for $n$ = 2. Minima of all the curves 
for a particular $n$ value is the same. }
%\caption{Flux energy relationship for a 8 spin up electron ring corresponding to m = -1 (dotted), 0 (solid), 1 (dashed) for n = 1 and 2. For n=2 we have subtracted 1.7 from the original value.}

\label{pn}
\end{figure}

Hence,
\begin{equation}
\Theta(\theta) = \frac{1}{\sqrt{2\pi}}exp\left(i ({\phi \over \phi_0} \pm m') 
\theta \right)
\label{phi}
\end{equation}
Let
\begin{eqnarray}
x' = \sqrt{\frac{2m^*E}{\hbar^2}}r  = k r 
\label{xm}
\end{eqnarray}
Using Eq. (\ref{xm}) in Eq. (\ref{r1}) one gets the Bessel's equation of first kind given by
\begin{eqnarray}
x'^2 \frac{d^2R}{dx'^2} + x'\frac{dR}{dx'} + \left(x'^2- {m'}^2 \right)R=0
\label{rad}
\end{eqnarray}
The solution of Eq. (\ref{rad}) is given by
\begin{eqnarray}
R(r) =  A_{m'} J_{m'} \left(k r\right) + B_{m'} N_{m'} \left(k r\right)
\label{R}
\end{eqnarray}
where $J_{m'}$ and $N_{m'}$ are the Bessel and Neumann function of order $m'$. The boundary conditions for the radial function $R(r_{in}) = R(r_{out}) = 0$ gives 
\begin{eqnarray}
&&  -\frac{A_{m'}}{B_{m'}} = \frac{N_{m'} \left(k r_{in}\right)}{J_{m'} \left(k r_{in}\right)} =  \frac{N_{m'} \left(k r_{out}\right)}{J_{m'} \left(k r_{out}\right)} \label{AmBm} \\
&& N_{m'} \left(k r_{in}\right) J_{m'} \left(k r_{out}\right) = N_{m'} \left(k r_{out}\right) J_{m'} \left(k r_{in}\right) \label{m'}
\end{eqnarray}
Eq. (\ref{m'}) determines the allowed values of energy for a particular $m'$. So the complete solution can now be written by combining 
%\begin{eqnarray}
%\psi\left(r,\theta \right) = \left[A_{m'} J_{m'} \left(k_n r\right) + B_{m'} N_{m'} \left(k_n r\right)\right] \frac{1}{\sqrt{2\pi}}exp\left(i m \theta \right)
%\label{psi-a}
%\end{eqnarray} 
Eqs. (\ref{R}) and (\ref{phi}) as,
\begin{eqnarray}
 \psi( r,\theta) &=&  \left[ \left(   A_{m'} J_{m'} \left(k_{n} r\right) + B_{m'} N_{m'} \left(k_{n} r\right) \right) \right. \nonumber \\
 & &\left. \left(  \frac{1}{\sqrt{2\pi}}exp\left(i (m'+\frac{\Phi}{\Phi_0})\theta \right) \right) \right]
  \label{psi-b}
\end{eqnarray}
$ k_n $ should be defined as $ E_{n}=\frac{\hbar^{2}k_{n}^2}{2m^*} $. 
Note that there seems to be no obvious decoupling of the radial part and 
the azimuthal part in $\Psi(r,\theta)$. The eigen energy can not be expressed 
as a sum of two terms and the total energy has to be found by 
solving Eq. (\ref{m'}). But if we solve Eq. (\ref{m'}) numerically and plot 
the eigen energies then the radial and the azimuthal part appears to get 
decoupled. This is shown for $n=1$, and $n=2$ in Fig. \ref{pn}. Possible $m'$ 
values corresponding to a particular $n$ value form a distinct band
signifying decoupling of radial energy and azimuthal energy. Each curve 
is a parabola for both $n=1$ and $n=2$ just like what we get in a one 
dimensional ring.

We can get further insight when we move into higher dimensions by noting the similarities 
between Q1D and 1D. Multiplying both sides of Eq. (\ref{r1}) and Eq. (\ref{p1}) by $\frac{\hbar^2}{2m^* r^2}$and adding $\frac{m^* \omega_c^2 r_f^2}{8 r^2}$ to both sides of Eq. (\ref{p1}) one can obtain the following equations
\begin{eqnarray}
-\frac{\hbar^2}{2m^*}\frac{1}{r}\frac{d}{dr} \left( r \frac{dR}{dr} \right) + \left(\frac{m^* \omega_c^2 r_f^2}{8 r^2} + \frac{\lambda \hbar^2}{2 m^* r^2} -E \right)R = 0
\label{r2} \\
\frac{1}{2m^*}\left( \hat{p}_\theta - \frac{e}{c} A(\vec{r}) \right)^2 \Theta(\theta) = \left(\frac{m^* \omega_c^2 r_f^2}{8 r^2} + \frac{\lambda \hbar^2}{2 m^* r^2}\right)\Theta
\label{p2}
\end{eqnarray}
where, $\hat{p_\theta}=-i\hbar\frac{1}{r}\frac{\partial}{\partial \theta}$.

If we write,
\begin{eqnarray}
E_1(r) = E - \frac{m^* \omega_c^2 r_f^2}{8 r^2} - \frac{\lambda \hbar^2}{2 m^* r^2} = E - \frac{\hbar^2 m'^2}{2m^*r^2}
\label{E1r1}
\end{eqnarray}

then Eq. (\ref{r2}) reduces to a simple form
\begin{equation}
-\frac{\hbar^2}{2m^*}\left[ \frac{1}{r}\frac{d}{dr} \left( r\frac{d}{dr}\right) \right] R(r) = E_1(r)R(r)
\label{E1r}
\end{equation}

Using Eq. (\ref{E1r}) in Eq. (\ref{p2}), we further obtain
\begin{eqnarray}
\frac{1}{2m^*}\left( -i\hbar \frac{1}{r}\frac{\partial}{\partial \theta} - \frac{e}{c} A(\vec{r}) \right)^2 \Theta(\theta) = \left( E - E_1(r) \right)\Theta(\theta)
\label{pe1}
\end{eqnarray} 

Since $rd\theta = dx$, Eq. (\ref{pe1}) becomes
\begin{eqnarray}
-\frac{\hbar^2}{2m^*} \left(\frac{d}{dx}-\frac{ie A(r)}{c\hbar} \right)^2 \Theta(x) = [ E - E_1(r) ]\Theta(x)
\label{c1d}
\end{eqnarray} 
$\Theta$ which is a function of $m'$ and $\theta$ in Eq. (\ref{pe1}) now becomes 
a function of $k$ and $x$ in Eq. (\ref{c1d}).

The $r$ dependence of $A$ is of no consequence as $A(r)$ in Eq. (\ref{c1d})
can be gauged away in a manner just as one does in the 1D case to give
%As $rd\theta = dx$ (\ref{pe2}) becomes
\begin{equation}
-\frac{\hbar^2}{2m^*} \frac{d^2}{dx^2} \Theta'(x)= [E - E_1(r)]\Theta'(x)
\label{pe2}
\end{equation}
where $ \Theta' $, the gauge transformed version of $ \Theta $ is given by
\begin{equation}
\Theta'(x) = \Theta(x) e^{-i\frac{e}{\hbar c}\int A(r) rd\theta }
\end{equation}

Ref. [\onlinecite{cheung}] writes for a one dimensional ring
\begin{equation}
-\frac{\hbar^2}{2m^*} \frac{d^2}{dx^2} \Theta'(x)= E_{1D}\Theta'(x)
\label{pe3}
\end{equation}

\begin{figure}
\includegraphics[height=4truecm,width=12truecm]{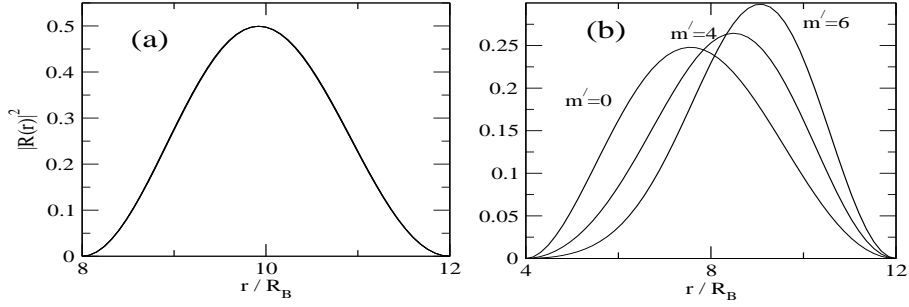}
\caption{Probability distribution of radial wave function of an electron 
in a ring with (a) $r_{in}=8R_B$, $r_{out}=12R_B$ and 
(b) $r_{in}=4R_B$, $r_{out}=12R_B$ for different $m'$ values. 
In (a) all the curves corresponding to different $m'$ values overlap on 
each other but in (b) they are distinct (here we have shown for 
$m'$=0, $m'$=4, $m'$=6). Situation in (a) leads to
a broken symmetry state and for (b) symmetry is restored.}
\label{radwf}
\end{figure}

Note that Eq. (\ref{pe2}) and Eq. (\ref{pe3}) give similar energy spectrum 
as has already been clearly explained in Fig. \ref{pn}. $E- E_1(r)$ 
in Eq. (\ref{pe2}) corresponds to $E_{1D}$ in Eq. (\ref{pe3}). Only difference is that 
$[E-E_1(r)]$ in Eq. (\ref{pe2}) has to be determined from Eq. (\ref{E1r}) 
and the radial wave function compensates in such a way that the eigen 
energies of the system turn out to be similar in quasi one dimension as well as 
in one dimension. This is true for a narrow ring as well as a wide 
ring. A gradual crossover to an extremely wide ring that can be 
considered as a two dimensional system does not change this feature. 
The only change of feature will be in the nature of the radial wave function. 
This is well demonstrated in Fig. \ref{radwf} where, the probability density 
is plotted across the radius of the ring for different $m'$ values 
corresponding to $n$=1. For narrow rings, the wave functions corresponding
to possible different $m'$ values look identical. However, for wider rings
this is not true ; for two different $ m' $ values there is a lot of difference 
in the wave function profile. We have plotted only for a few values of $m'$ 
just for the sake of visual clarity of the figures.

\begin{figure}[h]
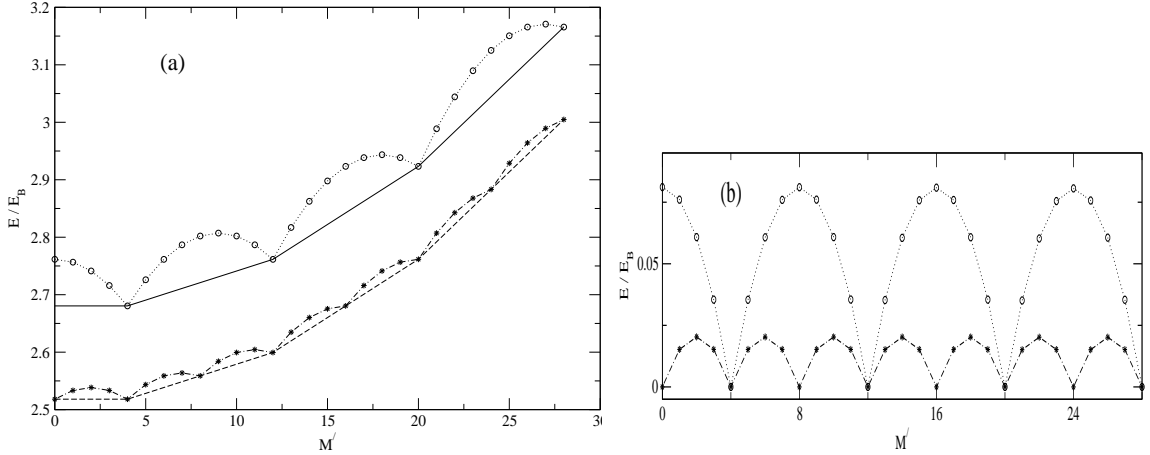

\includegraphics[height=6truecm,width=8truecm]{2d1.eps}
\includegraphics[height=4truecm,width=7truecm]{2d2.eps}
\caption{(a) Yrast spectra for electrons in a Q1D ring of inner 
radius $r_{in}=8R_B$ and outer radius $r_{out}=12R_B$. 
The dotted line gives the yrast spectra for 8 spin up electrons 
(calculated data plotted as circles). 
The solid line is drawn by connecting the 
local minima of the dotted line. The dash-dot line is the yrast 
spectra for 4 spin up and 4 spin down electrons
(calculated data plotted as stars). 
The dashed line is drawn connecting the local minima of the dash-dot 
line. In both cases the electrons are interacting only through 
statistics and Coulomb interaction is ignored.
(b) dotted line is obtained by subtracting the solid line from the dotted 
line in (a) and the dash-dot line is obtained by subtracting the dashed 
line from the dash-dot line in (a).}
\label{2d1}
\end{figure}

\subsection{Effect of Fermi Statistics}
In Fig. \ref{2d1}a we consider the same parameters as are used in Fig. \ref{pn}a 
and plot the yrast spectra for (i) 8 up spin electrons(plotted as circles)
and (ii) 4 up, 4 down 
(plotted as stars) spin electrons. Note that, we are not including the 
effect of Coulomb interaction but consider the consequences of Fermi 
statistics only. The dash-dot line and the dotted line 
are guides to the eye. It can be seen clearly that,
we obtain an identical behaviour in quasi 
one dimension (Fig. \ref{2d1}a) as compared to in one dimension
depicted in Fig. \ref{1d}a. This demonstratively 
signifies decoupling of the center of mass energy and the energy associated 
with the internal degrees of freedom. This in turn, implies the breakdown of 
symmetry in the internal frame as has already been explained along with Fig. \ref{1d}a. 
We see that, the local minima increases parabolically with the magnitude of 
flux exactly as it happens in a 
one dimensional ring. We fit the local minima to $M'^2/2I$, where $M'$ is 
the designated total angular momentum which has been calculated quantum 
mechanically i.e. using the relation,
$M'=\Sigma m'_i$. We have used $I$, the moment of inertia
as a fitting parameter and have obtained the value of $I$ to 
be 791.4 $m_eR_B^2$ for a ring of inner radius 8$R_B$ and outer radius 
12$R_B$. The moment of inertia for 8 classical electrons placed at 
equal distances in a ring like arrangement and rotating on a ring of 
radius 10$R_B$ is 800 $m_eR_B^2$. 
This further confirms a semi rigid classical structure and hence symmetry 
breaking. In Fig. \ref{2d1}b, where like in Fig. \ref{1d}b, we have used the 
same procedure of subtracting the original data from the parabolic line. 
The perfect periodicity of a broken symmetry state is again clear from 
the periodicity of the current (modulo 8 and 4).

\begin{figure}
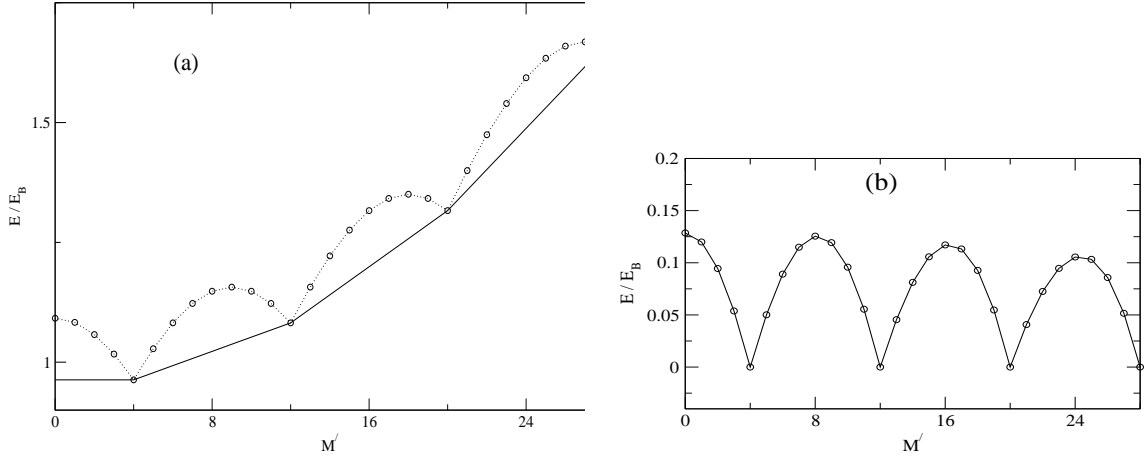

\includegraphics[height=6truecm,width=8truecm]{412.eps}
\includegraphics[height=4truecm,width=7truecm]{thickness.eps}
\caption{Change in periodicity with variation of thickness for a ring 
with 8 spin up electrons. Here $r_{in} = 4R_B$ and $r_{out} =12R_B$. 
(a) Yrast spectra of the ring. Circles are the exact data points which 
are connected by dotted line and the minima are connected by a solid line. 
(b) Energy spectra after subtracting the solid line from the dotted 
line in (a). The periodicity is broken unlike in Fig. \ref{2d1}.}
\label{thickness}
\end{figure}

We will now show that in quasi one dimension unlike in one dimension there can be a transition that can be effected by increasing the thickness of the ring. Hence we plot the yrast spectra (Fig. \ref{thickness}a) for a ring of inner radius 
4$R_B$ and outer radius 12$R_B$.
We subtract the parabolic data from the 
original data and plot the resulting value. It is not exactly periodic
as can be seen in Fig. \ref{thickness}b. The periodicity is destroyed for larger 
$M'$ (Fig. \ref{thickness}b). 
We again fit the local minima to
$M'^2/2I$ using $I$ as fitting parameter. The value of $I$ we obtain from
the fitting is 551.8 $m_eR_B^2$. The moment of inertia for 8 classical
electrons sitting at equal distance in a ring like arrangement and rotating on a circle of radius
8$R_B$ is 512.0 $m_eR_B^2$. Hence in this case we can't say that the
particles are behaving almost as classical particles which are localized at
equal distances in the
center of mass frame. 
In one dimension, statistics plays a major role as the particles can not 
cross each other. In a two dimensional ring with inner radius 8$R_B$ 
and outer radius 12$R_B$, we see that the radial probability distribution 
for all $m'$ values coincide with each other (Fig. \ref{radwf}a). In this case,
it looks like the effect is similar to the case of one dimension and that the particles 
can not cross each other. For a ring of inner radius 4$R_B$ and outer radius 
12$R_B$ the position of peaks of the radial probability distribution 
changes with $m'$ (Fig. \ref{radwf}b). In this case it appears that unlike in
1D the particles can cross each other inside the ring.  

\begin{figure}
\includegraphics[height=6truecm,width=10truecm]{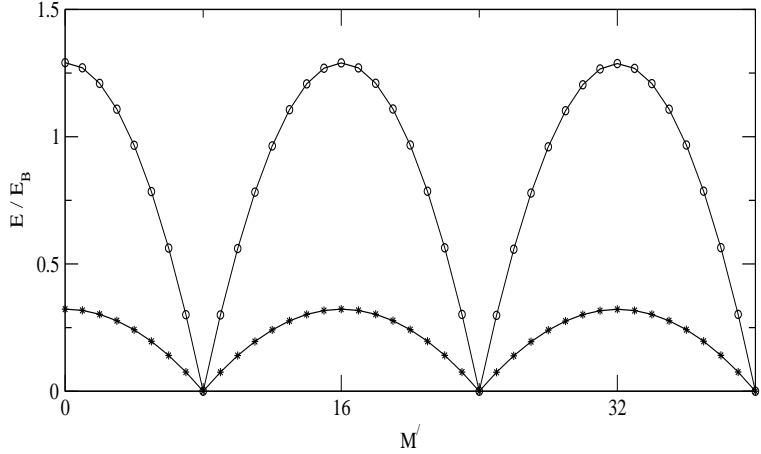}
\caption{Periodicity is restored when the radius and thickness are changed simultaneously. Stars are data for $r_{in} = 4R_B$ and $r_{out} =6R_B$, and circles are data for $r_{in} = 8R_B$ and $r_{out} =12R_B$ for a ring with 16 spin up electrons.}
\label{rthickness}
\end{figure}

From Fig. \ref{thickness} and associated narrations, it seems that 
symmetry breaking is not possible in two dimensions that can be obtained 
by gradually increasing the thickness of a Q1D ring. 
But if we increase the radius and the thickness simultaneously, it results in
a symmetry breaking as we will explain now.
We take 16 electrons interacting only due to Fermi statistics and 
compare the energy spectra for two rings. One with inner radius 4$R_B$ and 
outer radius 6$R_B$ and another with inner radius 8$R_B$ and outer 
radius 12$R_B$. Here we have not plotted the yrast spectra, but plotted 
the energy values after subtracting the parabolic data we obtain by 
connecting the minima (Fig. \ref{rthickness}). We see that both the 
curves show the same periodicity signifying broken symmetry. 
The amplitude of energy for the ring with inner radius 8$R_B$ and outer 
radius 12$R_B$ is 1/4-th of the amplitude for the ring with inner 
radius 4$R_B$ and outer radius 6$R_B$. This is also similar to the case 
of a one dimensional ring, where if one makes the radius double, the 
energy becomes 1/4-th. Hence it seems quite natural that, by increasing
the radius and the 
thickness simultaneously one can observe symmetry breaking for an 
infinite two dimensional system as well. This is because the act of 
increasing
the radius consequently lowers the azimuthal component of the momentum  
and the act of increasing the thickness lowers the 
radial component of the momentum.
Thus it seems that even in 2D, one can find very low energy states
with a broken symmetry.

\subsection{ Effect of Coulomb Interaction}

Wave function for the Hamiltonian in Eq. (\ref{h}) can be written in terms
of $\psi(r, \theta)$ of Eq. (\ref{psi-b}) as the many particle extension
\begin{eqnarray}
\Psi_N(\left\lbrace r_j,\theta_j \right\rbrace) &=& \sum_{\left\lbrace m'_j \right\rbrace} a_{\left\lbrace m'_j \right\rbrace} \mathcal{A} \prod_{j=1}^N \left[ \left(   A_{m'_j} J_{m'_j} \left(k_{n_j} r_j\right) + B_{m'_j} N_{m'_j} \left(k_{n_j} r_j\right) \right) \right. \nonumber \\
& &\left. \left( \frac{1}{\sqrt{2\pi}} e^{i (m'_j + \Phi / \Phi_0) \theta_j} \right) \right]
\label{psi0}
\end{eqnarray} 
where $a_{\left\lbrace m'_j \right\rbrace}$s are unknown coefficients.   
$ \left\lbrace m'_j \right\rbrace $ corresponds to the set of
all allowed $ m'_j $ values.

We now introduce the center of mass ($\xi$) and relative 
($\zeta_i$) coordinates defined as
 \begin{eqnarray}
 \xi = \frac{1}{N} \sum_{j=1}^N \theta_j \label{cmc} \\
 \zeta_j = \theta_j - \xi  \label{rc} 
 \end{eqnarray}
 
 The many body Hamiltonian (Eq. (\ref{h})) in terms of the center of mass and the relative coordinates is given by
 \begin{eqnarray}
 \bf{H} &=& \sum_{j=1}^N\frac{-\hbar^2}{2m^* r_j^2} \frac{1}{N^2} \frac{\partial^2}{\partial \xi^2} 
+ \sum_{j=1}^N \left( \frac{-\hbar^2}{2m^*} \frac{1}{r_j} \frac{\partial}
{\partial r_j} \left( r_j \frac{\partial}{\partial r_j} \right) +
V(r_j) \right) \nonumber \\
&&  + \sum_{j=1}^N\frac{-\hbar^2}{2m^* r_j^2} \frac{\partial^2}{\partial \zeta_j^2}
  + \sum_{j=1}^N\frac{-\hbar^2}{2m^* r_j^2}\frac{1}{N^2} \left( \sum_{k=1}^N  \frac{\partial}{\partial \zeta_k} \right)^2  \nonumber \\
&& +\sum_{j=1}^N \frac{2}{N}\frac{-\hbar^2}{2m^*r_j^2}\frac{\partial}{\partial \xi} \frac{\partial}{\partial \zeta_j}
-\sum_{j=1}^N \frac{2}{N}\frac{-\hbar^2}{2m^*r_j^2}\frac{\partial}
{\partial \xi} \sum_{k=1}^N \frac{\partial}{\partial \zeta_k}
-\sum_{j=1}^N \frac{2}{N}\frac{-\hbar^2}{2m^*r_j^2}\frac{\partial}
{\partial \zeta_j} \sum_{k=1}^N \frac{\partial}{\partial \zeta_k} \nonumber \\
&& +\frac{1}{2}\sum_{i \neq j} \frac{1}{4\pi \epsilon} \frac{e^2}{\sqrt{r_i^2 + r_j^2 - 2r_i r_j \cos(\zeta_i - \zeta_j)}}
\label{MH}
\end{eqnarray}
Here, $N$ is the total number of electrons.

Our non-interacting calculations suggests that in some regime if 
$r_j$ = constant = $r_{1D}$ then the first term in Eq. (\ref{MH})is 
$\hat{M'}^2 / 2I$ where $\hat{M'}=-i \hbar \frac{\partial}{\partial \xi}$ 
and $I= m^*Nr_{1D}^2$. Note $-i\hbar \frac{\partial}{\partial \xi}$ 
commutes with $ \textbf{H} $ in Eq. (\ref{MH}). This essentially implies that 
Coulomb interaction (the last term) 
will not change the center of mass angular 
momentum and hence $\sum m'_j = M'$ (where $M'$ is the eigenvalue 
corresponding to $\hat{M'}$) will be a conserved quantity. Its value remains the same whether the last term in Eq. (\ref{MH}) is included or not. 
Therefore, substituting for $\theta_j$ from Eq. (\ref{cmc}) and
Eq. (\ref{rc}) in Eq. (\ref{psi0})

\begin{eqnarray}
  \Psi(\left\lbrace r_j,\theta_j \right\rbrace) &=& e^{i\left( M'+N \frac{\Phi}{\Phi_0} \right)\xi} \sum_{\left\lbrace m'_j \right\rbrace} a_{\left\lbrace m'_j \right\rbrace} \mathcal{A} \prod_{j=1}^N \left[ \left(   A_{m'_j} J_{m'_j} \left(k_{n_j} r_j\right) + B_{m'_j} N_{m'_j} \left(k_{n_j} r_j\right) \right) \right. \nonumber \\
 & &\left. \left(  \frac{1}{\sqrt{2\pi}} e^{i m'_j \zeta_j} \right) \right]
  \label{psi2}
\end{eqnarray} 
This is because if we switch off the Coulomb interaction then the sum 
will not appear in the wave function and hence, the exact wave function is
\begin{eqnarray}
  \Psi(\left\lbrace r_j,\theta_j \right\rbrace) &=& e^{i\left( M'+N 
\frac{\Phi}{\Phi_0} \right)\xi} \mathcal{A} \prod_{j=1}^N 
\left[ \left(   A_{m'_j} J_{m'_j} \left(k_{n_j} r_j\right) + 
B_{m'_j} N_{m'_j} \left(k_{n_j} r_j\right) \right)  
\left( \frac{1}{\sqrt{2\pi}} e^{i m'_j \zeta_j} \right) \right]
\label{psi3}
\end{eqnarray} 
This argument can be given for each term in the sum of
Eq. (\ref{psi2}). So note that the flux dependence remains the 
same in presence or absence of Coulomb interaction. 
This is because the term inside the summation in Eq. (\ref{psi2}) does not 
contain flux and the periodicity will be $\Phi_0/N$ as can be seen from Eq. (\ref{psi3})
provided we start from the periodic structure as in Fig. \ref{2d1}.
If we do not start from a periodic structure then the $N \Phi/\Phi_0$ in
the exponent of Eq. (\ref{psi2}) does not imply a $\Phi_0/N$ periodicity. 
Aharonov-Bohm effect in a ring can be observed \cite{tapash}. So changes in periodicity can also be observed and can give us the demonstration of
symmetry breaking transition. 

\section{Three dimensions}
\begin{figure}[h]
\includegraphics[width=8truecm,height=4truecm]{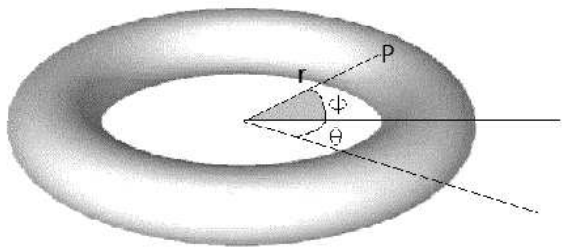}
\caption{A three dimensional ring}
\label{3dring}
\end{figure}

For a three dimensional ring (shown in Fig. \ref{3dring}), the single particle
Schr\"{o}dinger equation is given by
\begin{equation}
\left( -\frac{\hbar^2}{2m^*} \left[ \frac{1}{r^2}\frac{\partial}{\partial r} \left( r^2 \frac{\partial}{\partial r} \right)+\frac{1}{r^2 sin\theta} \frac{\partial}{\partial \theta} \left( sin \theta \frac{\partial}{\partial \theta} \right) + \frac{1}{r^2 sin^2 \theta} \frac{\partial^2}{\partial \phi^2}\right] + V(r,\theta ,\phi) \right) \Psi = E \Psi
\label{sc1}
\end{equation}
where the potential is defined as 
\begin{eqnarray}
V(r, \theta, \phi) &=& 0 \hspace{1cm} inside~ the~ shaded~ region 
\nonumber \\
     &=& \infty \hspace{1cm} everywhere~ else
\end{eqnarray}

If we express the total  as $\Psi = R(r)P(\theta, \phi)$ then Eq. (\ref{sc1}) becomes
\begin{equation}
-\frac{1}{R} \left[ \frac{\partial}{\partial r} \left( r^2 \frac{\partial}{\partial r} \right) - \frac{2m^*{r^2}}{\hbar^2}V(r) + \frac{2m^*{r^2}E}{\hbar^2} \right] R(r) = \frac{1}{P} \left[ \frac{1}{sin\theta} \frac{\partial}{\partial \theta} \left( sin \theta \frac{\partial}{\partial \theta} \right) + \frac{1}{ sin^2 \theta} \frac{\partial^2}{\partial \phi^2} \right]P(\theta, \phi)
\label{sc2}
\end{equation}

If $\theta$ is very small then $sin\theta = \theta$ and Eq. (\ref{sc2}) becomes
\begin{equation}
-\frac{1}{R} \left[ \frac{\partial}{\partial r} \left( r^2 \frac{\partial}{\partial r} \right) - \frac{2m^*{r^2}}{\hbar^2} V(r) + \frac{2m^*{r^2}E}{\hbar^2} \right] R(r) = \frac{1}{P} \left[ \frac{1}{\theta} \frac{\partial}{\partial \theta} \left( \theta \frac{\partial}{\partial \theta} \right) + \frac{1}{\theta^2} \frac{\partial^2}{\partial \phi^2} \right]P(\theta,\phi)=\lambda
\label{sc3}
\end{equation}

Once again the radial part effectively decouples and one can have
\begin{equation}
\left[ \frac{1}{\theta} \frac{\partial}{\partial \theta} \left( \theta \frac{\partial}{\partial \theta} \right) + \frac{1}{\theta^2} \frac{\partial^2}{\partial \phi^2}  \right]P(\theta,\phi)= \lambda P(\theta,\phi)
\label{sc4}
\end{equation}
If we replace $\theta$ by $r$, then Eq. (\ref{sc4}) will look identical to
Eq. (\ref{sch1}) in absence of magnetic field. Magnetic field and Coulomb 
interaction can then be treated in the same way as we have done for 
Eq. \ref{sch1}. 
In such a ring therefore the equation of motion is just like in 1D and hence
we can again expect symmetry breaking. However, if $sin\theta$ can not be 
approximated by $\theta$ then we should not expect symmetry breaking in 
three dimensions. Thus we cannot get symmetry breaking in atoms.

\section{conclusions}

We show that in quasi-one-dimension and two dimensions, owing to 
internal symmetry
breaking transition can take place for a many electron state. 
In the broken symmetry state, the electrons crystallize in the
internal frame and behave like a semi-rigid rotor. The low-lying
excitations are associated with rotations and vibrations of this
semi-rigid rotor.
While it is known that one-dimensional systems always show a broken 
symmetric state and no transition, the broken symmetry state in Q1D is 
identical
to that in 1D and the transition is unique to Q1D and 2D. In three
dimensions however, it is unlikely that a broken symmetric state can exist.
One can experimentally verify the broken symmetry state by taking a
ring made up of the Q1D wire and generating a persistent current
in the ring. The flux periodicity of the persistent current
gives the signature of a broken symmetry state. When the system becomes
like a semi-rigid rotor, the flux periodicity becomes $\phi_0/N$,
where $N$ is the number of electrons in the ring. In the symmetric state,
the $\phi_0/N$ periodicity is destroyed.
In finite systems, the transition is always gradual as is 
generally expected.

%\section*{ACKNOWLEDGEMENT}
\section{acknowledgement}
We would like to thank Prof. M. Manninen for useful discussions.

\end{document}